%% file: DileptonSigmaPLB.tex
\documentclass[twocolumn,10pt,5p]{elsarticle}

\usepackage{graphicx}

\usepackage{amssymb,amssymb}

\usepackage{atlasphysics}
\usepackage{subfigure}
\usepackage{mathrsfs}
\usepackage{slashed}
\usepackage{setspace}
\usepackage{amsmath}
\usepackage{multirow}
\usepackage{hyperref}

\newlength{\figwidth}
\newcommand{\pmasym}[2]{^{+#1}_{-#2}}
\newcommand{\sigmattbar}{\ensuremath{\sigma_{\ttbar}}}

\newcommand{\lumitot}{\mbox{$35~\mathrm{pb}^{-1}$}}
\newcommand{\xsectot}{177}
\newcommand{\xsecstat}{\pm 20}
\newcommand{\xsecsyst}{\pm 14}
\newcommand{\xseclumi}{\pm 7}

\newcommand{\xsecbtot}{194}
\newcommand{\xsecbstat}{\pm 23}
\newcommand{\xsecbsyst}{\pmasym {18}{14}}
\newcommand{\xsecblumi}{\pm 7}
\newcommand{\AKT}{anti-k$_{t}$}

\def\MCatNLO{{\sc MC@NLO}}

\def\JetProb{{\sc JetProb}}

\def\Zg{\Zboson/\gamma^{*}}
\def\GEANT{{\sc Geant4}}

\begin{document}

\begin{frontmatter}  

  \title{\vspace{-0.5cm} 
  \flushright{\normalsize{CERN-PH-EP-2011-103}} \\ 
  \center{Measurement of the Top Quark Pair Production Cross Section in $pp$\  Collisions at $\sqrt{s}=7$~TeV in Dilepton Final States with ATLAS}
  }

\journal{Phys. Lett. B}
\author{The ATLAS Collaboration}

\date{19 June 2011}

\begin{abstract}
A measurement of the production cross section of top quark pairs
($\ttbar$) in proton-proton collisions at a center-of-mass energy of $7\TeV$ 
recorded with the ATLAS detector at the Large Hadron Collider is
reported.  Candidate events are
selected in the dilepton topology with large missing transverse energy
and at least two jets.
Using a data sample corresponding to an integrated luminosity of \lumitot, a \ttbar\ production cross section 
$\sigmattbar=\xsectot \xsecstat ({\rm stat.})\xsecsyst ({\rm syst.}) \xseclumi ({\rm lum.})~{\rm pb}$
is measured for an assumed top quark mass of $m_t=172.5\GeV$.
A second measurement requiring at least one jet identified as coming from 
a $b$~quark yields a comparable result, demonstrating that the 
dilepton final states are consistent with being accompanied by $b$-quark jets.
These measurements are in good agreement with Standard Model predictions.
\end{abstract}

\begin{keyword}
top physics, heavy quark production, total cross section
\end{keyword}

\end{frontmatter}  

\setcounter{page}{1}

\section{Introduction}\label{s:intro}

The study of  top quarks 
probes the validity of the Standard Model~(SM)
and plays an important role in the search for new physics. 
At the Large Hadron Collider (LHC)  
the $\ttbar$ production cross section  (\sigmattbar) in proton-proton ($pp$) collisions at a center-of-mass energy
$\sqrt{s}=7 \TeV$ is predicted by an 
approximate next-to-next-to-leading-order (NNLO) SM calculation to be
$165\pmasym{11}{16}$~pb~\cite{topxs,Langenfeld:2009tc}.
A measurement of \sigmattbar\ in
various decay channels tests perturbative QCD and the
description of top quark decay. 
Moreover, $\ttbar$\ production is an important background
in searches for the Higgs boson and physics beyond the Standard Model. 
The study of $\ttbar$\ events may provide 
evidence for new physics that modifies the production and/or decay of
top quarks.

In the SM the top quark decays to a $\Wboson$ boson 
and a $b$~quark
($t\rightarrow Wb$) with a branching ratio close to 100\%~\cite{SM,Glashow1961579,Salam:1968}.
The $\ttbar$ event
topologies are determined by the decays of the two $W$ bosons: a pair
of quarks ($\Wboson \rightarrow q q'$) or a
lepton-neutrino pair ($\Wboson\rightarrow\ell\nu_\ell$), where
$\ell$ refers to an electron, muon or tau lepton and $\nu_\ell$ is the corresponding neutrino. 
Top quark production in dilepton final states has been previously studied using
proton-antiproton collisions at
$\sqrt{s}=1.96~\TeV$~\cite{Aaltonen:2010bs,Abazov:2011cq} and
LHC measurements have recently been reported in several
final states~\cite{Chatrchyan:2011nb,top2010}.
In this letter, we present a measurement of the \ttbar\ production cross section using the dilepton channel,
in which both $\Wboson$\ bosons decay to leptons. 
A selected event should exhibit two opposite-sign leptons, unbalanced
transverse momentum indicating the presence of neutrinos from the $W$-boson
decays and two $b$-quark jets.
The measurement is performed with ten times more data than the previous ATLAS observation
of $\ttbar$ production~\cite{top2010}.

The \ttbar~dilepton final states can be
efficiently selected using 
kinematic requirements on the final state objects.
To further reduce backgrounds and verify that the dilepton final states
are accompanied by $b$-quark jets,
a separate measurement is performed
requiring the presence of a jet 
identified as coming from a $b$~quark and relaxing the 
kinematic selection.
Both cross section measurements are reported in this letter.
Leptons are either well-identified electron or muon candidates or, to reduce losses from lepton identification inefficiencies, 
isolated tracks (referred to as track-lepton candidates). 
Selected events have either two well-identified lepton candidates ($ee$, $\mu\mu$ and $e\mu$), or one
well-identified lepton candidate and one track-lepton candidate ($e$TL and $\mu$TL),
together creating five separate dilepton channels.
Each selected dilepton channel is exclusive, i.e.\ has no overlap with the other channels.
Tau leptons are not explicitly reconstructed, but reconstructed leptons can arise from leptonic tau decays and a track-lepton can arise from all one-prong tau decay modes.

The number of candidate events in the selected sample 
is corrected for background contributions from $\Zg$+jets, single top and
diboson production, and from events with misidentified lepton candidates.
The cross section is measured taking into account the $\ttbar$\ signal 
acceptance. 
The primary background contributions are estimated using complementary
data samples to reduce the uncertainties
associated with the simulation and theoretical calculations of background rates.

\section{Detector and Data Sample}
\label{s:detector}

The ATLAS detector~\cite{atlasdet} at the LHC covers nearly the entire
solid angle\footnote{ATLAS uses a right-handed coordinate system with its origin at the nominal interaction point (IP) in the center of the detector and the $z$-axis along the beam pipe. 
The $x$-axis points from the IP to the center of the LHC ring, and the $y$ axis points upward. 
Cylindrical coordinates $(r,\phi)$ are used in the transverse plane, $\phi$ being the azimuthal angle around the beam pipe. 
The pseudorapidity is defined in terms of the polar angle $\theta$ as $\eta=-\ln\tan(\theta/2)$. Distances in $\eta-\phi$ space are given as $\Delta R = \sqrt{\phi^2+\eta^2}$.}\ around the collision point. 
It consists of an inner tracking detector 
(ID) comprising a silicon pixel detector, a silicon microstrip detector 
(SCT), and a transition radiation tracker, providing tracking capability within $|\eta|<2.5$. 
The ID is surrounded by a thin superconducting solenoid providing a 2~T 
magnetic field, and by liquid-argon (LAr) electromagnetic sampling calorimeters with high granularity.
An iron-scintillator tile calorimeter 
provides hadronic energy measurements in the central rapidity 
range ($|\eta|< 1.7$). 
The end-cap and forward regions are instrumented with LAr calorimetry for 
both electromagnetic and hadronic energy measurements up to $|\eta|<4.9$.
The calorimeter system is surrounded by a muon spectrometer 
incorporating three 
superconducting toroid magnet assemblies.

A three-level trigger system is used to select the high-$\pt$\ events
for this analysis. 
The level-1 trigger is implemented in hardware and uses a subset of
the detector information to reduce the rate to a design value of at most
75\,kHz. This is followed by two software based trigger levels,
that together reduce the event rate to
about 200\,Hz.
The analyses use collision data with a center-of-mass energy of
$\sqrt{s}= 7\TeV$\ recorded in 2010 
with an integrated luminosity of $35.3\pm1.2$~${\rm pb}^{-1}$~\cite{lumi}.

\section{Simulated Samples}
\label{s:mc}

Monte-Carlo (MC) simulation samples are used to calculate the $\ttbar$\ acceptance and 
to evaluate
the contributions from those background processes that are difficult to
estimate from complementary data samples.
All MC samples are processed with the
\GEANT~\cite{geant4}\ simulation of the ATLAS detector~\cite{:2010wqa}
and events are passed through the same analysis chain as the data.

The generation of \ttbar{} and single top quark events 
uses the \MCatNLO\  
generator~\cite{mcatnlo1,Frixione:2003ei,Frixione:2005vw}\ with the 
CTEQ6.6~\cite{cteq66}\ 
parton distribution function (PDF) set 
and a top quark mass of $172.5\GeV$.
The $\ttbar$\ cross section is normalized to the prediction
of {\sc Hathor}~\cite{Aliev:2010} that employs an 
approximate NNLO pertubative QCD calculation.  
Single top quark production with \MCatNLO\ includes the $s$, $t$ and $Wt$
channels, and the diagram-removal scheme~\cite{diagrem}\  is used
to reduce overlap with the $\ttbar$\ final state.

Drell-Yan events
($\Zg$+jets) are modeled with the {\sc Alpgen} 
generator using the MLM matching scheme~\cite{alpgen}\ 
and the CTEQ6L1 \cite{cteq6l}\ PDF set.  
The $\Zg$+jets
samples, including light and heavy flavor jets,
are normalized to NNLO calculations from the FEWZ program~\cite{2004PhRvD..69i4008A} with a $K$-factor of 1.25.
Background contributions from the $W$+jets final states come primarily
from events where the $W$\ boson decays leptonically and the second
lepton candidate is a misidentified jet. They are estimated using 
auxiliary data samples. 
All MC simulated events are hadronized using the {\sc Herwig}\ shower model~\cite{herwig1,Corcella:2002jc}\ supplemented by the
{\sc Jimmy}\ underlying event model~\cite{jimmy}.
Both hadronization programs are tuned to data using the ATLAS MC10 tune~\cite{ATLAS:1303025}.
Diboson $WW$, $WZ$\ and
$ZZ$\ events are modeled using the {\sc Alpgen}\ generator normalized with
$K$-factors of 1.26 ($WW$), 1.28 ($WZ$) and 1.30 ($ZZ$) to match the total cross 
section from NLO QCD
predictions using calculations with the MCFM program~\cite{Campbell:1999ah}. 

For backgrounds, such as $W$+jets and QCD multijet events, that are mainly selected through non-prompt or misidentified leptons, simulated MC samples are not used, but instead data-driven estimations are employed (see Section~\ref{s:backgrounds}).

\section{Object Selection}
\label{s:object}

Electron candidates are reconstructed from energy deposits
in the  calorimeter, which are then associated to
reconstructed tracks of charged particles in the ID.
The candidates are required to pass a stringent 
selection~\cite{ATLAS:Tightelectrons}, which uses
calorimeter and tracking variables, and are required to have
\hbox{$\pt>20\GeV$} and
$|\eta|<2.47$. Electrons in the transition region
between the barrel and endcap calorimeters, defined as
$1.37<|\eta|<1.52$, are excluded.
%The variable $\eta$\ is the
%pseudorapidity of the calorimeter energy cluster associated with the
%candidate. 
%This selection provides the cleanest sample of electron candidates with 
%efficiency in excess of X.
%This selection places requirements on the ratio of the electron cluster energy
%measured in the calorimeter to the momentum in the tracker, 
%on an associated track hit in
%the innermost pixel layer when it passes through an active region
%and on high-threshold TRT hits when available.

Muon candidates are reconstructed~\cite{ATLAS-CONF-2011-063} by searching for track segments in
different layers of the muon chambers.
These segments are combined starting from the
outermost layer and matched
with tracks found in the ID. 
The candidates are refit using the complete track information from 
both detector
systems and are required to satisfy $\pt>20\GeV$ and
%systems~\cite{CSCbook}, and required to satisfy $\pt>20$~GeV and
$|\eta|<2.5$.

Both lepton candidates are required to be isolated to reduce backgrounds 
arising from jets
and to suppress the selection of leptons from heavy flavor
decays inside jets.
For electron candidates, the transverse energy ($\ET$) deposited in the calorimeter
and not associated to the electron is summed in a cone in $\eta - \phi$ space of radius $\Delta R = 0.2$
around the electron.
This $\ET$ is
required 
to be less than $4\GeV$. For muon candidates, both 
the corresponding calorimeter isolation $\ET$ and the analogous track isolation transverse momentum ($\pt$)
must be less than $4\GeV$ in a cone of $\Delta R = 0.3$.
The track isolation $\pt$ is calculated from the sum of
the track transverse momenta for tracks with $\pt>1\GeV$ around the muon candidate.
Additionally, muon candidates must
be separated by a distance $\Delta R > 0.4$\ from any jet with
$\pt>20\GeV$, further suppressing muon candidates from heavy flavor decays.

Muon candidates arising from cosmic rays are rejected by removing candidate 
pairs that are back-to-back in the $r-\phi$ plane and that have transverse 
impact parameter relative to the 
beam axis $|d_0| > 0.5$~mm.

Track-lepton candidates are defined by an ID track with $\pt > 20\GeV$
and a series of quality cuts optimized for high efficiency and 
discrimination between signal and the main background (non-$Z$ boson background, see Section~\ref{s:backgrounds}).
%The backgrounds are estimated from the observed rate of 
%candidates arising from misidentified track-leptons.
Tracks must have at least six  SCT hits  and at least one hit in the innermost pixel layer.
They also must have $|d_0| < 0.2~\mathrm{mm}$, and the uncertainty on the momentum measurement must be less than 20\%.
Tracks have to be isolated from other nearby tracks: the track isolation as
defined for muon candidates, but using tracks with $\pt>0.5\GeV$, must be less than $2\GeV$.
The use of track-lepton candidates primarily recovers acceptance losses from 
uninstrumented regions in the muon system and calorimeter transition regions.

Jets are reconstructed with the {\it \AKT} algorithm~\cite{antikt} with radius parameter $R = 0.4$
starting from energy clusters of adjacent calorimeter cells.
These jets are calibrated by first correcting the jet energy using the scale established
for electromagnetic objects and 
then performing a further correction to the hadronic energy scale using
$\pt$- and $\eta$-dependent correction factors obtained from
simulation~\cite{dijets}. 
Jets are corrected for additional energy deposits from the presence of multiple $pp$ interactions.
% and the position is corrected back to the position of the primary interaction (primary vertex).
The jets used in the analysis are required to have no
electron candidate or, in case of lepton+track events (see Sec.~\ref{s:event}),
no track-lepton candidate within $\Delta R=0.4$, 
$\pt>20\GeV$ 
and $|\eta|<2.5$.

Jets are identified as $b$-quark candidates using  
the \JetProb\ $b$-tagging algorithm~\cite{bib:JetProb}.
This algorithm takes all well-measured tracks associated with a given jet 
and forms a $p$-value\footnote{Probability value for a jet formed by the individual track probabilities.} for the hypothesis that the set of tracks comes from a
common primary vertex of a $pp$\ interaction, taking into account the track measurement
uncertainties.
The $p$-value requirement results in a $b$-tagging efficiency of $\approx 70\%$\ per jet in \ttbar~candidate events, and a mistag rate of order $1\%$ for both light-quark and gluon jets. 

The missing transverse energy ($\MET$) calculation begins with the vector 
sum of transverse momenta  
of all jets with $\pt>20\GeV$ and $|\eta| < 4.5$.
The transverse energies of electron candidates are added.
The contributions from all well-identified muon candidates and 
calorimeter clusters not belonging to a reconstructed object are also included.
To suppress backgrounds from $\Zg$+jets, the \MET\ is corrected by 
the \pt\ of the 
track-lepton in muon+track events if the $\Delta \phi$ between the \MET\ and track direction 
is less than 0.15 and there is no muon candidate within $\Delta R = 0.05$ of the track-lepton candidate. 
This properly accounts for the contribution to \MET\ of track-lepton candidates.

\section{Event Selection}
\label{s:event}

The analysis requires events selected online by an inclusive
single-lepton trigger ($e$ or $\mu$). 
The detailed trigger
requirements vary through the data-taking period, due to the rapidly
increasing LHC luminosity and the commissioning of the trigger system, but 
with a trigger threshold that ensures full efficiency for the lepton 
candidates with $\pt>20\GeV$ that are used in the analysis.
To ensure that the event was triggered by the selected lepton candidates,
one of the well-identified leptons and the trigger
object are required to match within $\Delta R < 0.15$.

Events are required to have a primary interaction vertex with at least five
tracks.  The event is discarded if any jet with $\pt>20\GeV$ fails 
quality cuts designed to reject jets arising from out-of-time activity
or calorimeter noise~\cite{jetqual}. 
If an electron candidate and a
muon candidate share a
track, the event is also discarded.

The selection of events in the signal region
consists of a series of kinematic requirements on the reconstructed
objects.
The requirements on
$\MET$, the dilepton invariant mass ($m_{\ell\ell}$), and the scalar $\pt$ sum of all selected jets and leptons ($\HT$) are
optimized using simulated events for maximum significance, defined as $S/\sqrt{S+{\sigma_B}^2}$\ 
where $S$\ is the expected number of signal events and ${\sigma_B}$\ is the
total uncertainty on the number of background events, $B$.

The presence of exactly two oppositely-charged 
well-identified lepton candidates is required.
If only one well-identified lepton candidate is found, the event is retained if an oppositely charged track-lepton candidate
is present, forming a lepton+track candidate event.
Events must have at least two jets with $\pT>20\GeV$ and $|\eta|<2.5$.
Furthermore, events in all channels other than $e\mu$\ are required to 
have $m_{\ell\ell} > 15\GeV$ in order to 
reject backgrounds from bottom quark production and vector meson decays.

The following additional kinematic requirements are made:
\begin{itemize}
\item Events in the $ee$ and $\mu\mu$ channels
  must satisfy $\MET>40\GeV$, and $m_{\ell\ell}$\ must differ
  by at least $10\GeV$ from the $Z$-boson mass, $m_Z$,
to suppress backgrounds from $\Zg$+jets and multi-jet events.
\item Events in the $e\mu$ channel have no \MET\ or $m_{\ell\ell}$\ cuts 
applied. 
In this case, remaining background from $\Zg$+jets production 
is suppressed by requiring $\HT>130\GeV$.
\item The lepton+track event candidates must have  $\MET>40\GeV$, \HT\ (including the 
track-lepton) $>150\GeV$, $|m_{\ell\ell} - m_Z| > 10\GeV$.
\end{itemize}

The requirement of at least one $b$-tagged jet using the
\JetProb\ algorithm allows for a kinematic event selection that can be 
optimized further.
To define the $b$-tagged sample, the selection described above is modified 
to require only events with two 
well-identified lepton candidates; the lepton+track candidates are discarded. 
The dilepton invariant mass must satisfy $|m_{\ell\ell} - m_Z| > 5\GeV$, and the 
$\MET$\ and $\HT$\ requirements are modified to $\MET>30\GeV$ and
$\HT> 110\GeV$.

The overall $\ttbar$\ signal efficiencies with respect to all \ttbar\ events (to all dilepton events) are
1.69\%\ (16.1\%) and 1.23\%\ (11.7\%) for the untagged and tagged analysis, respectively.

\section{Backgrounds}
\label{s:backgrounds}

The $\ttbar$ event selection is designed to reject $\Zg$+jets events.
However, a small fraction of such events 
will remain in the signal
sample primarily due to $\met$ mismeasurements. These events are difficult to 
model properly in simulations
due to large uncertainties on the non-Gaussian tails of the $\met$\ 
distribution, on the $Z$~boson cross section for higher jet 
multiplicities and on the lepton energy resolution. 
To estimate the $\Zg$+jets background (the $Z \to \tau\tau$\ channel is 
not considered here) in a data-assisted way, the number of $\Zg$+jets events is
measured in a control region orthogonal to the \ttbar\ dilepton signal region.
The control region is formed by events with the same jet requirements as the 
signal region, but with $|m_{\ell \ell} - m_Z| < 10\GeV$ and a lower $\MET$ cut 
($\met>15\GeV$ for the lepton+track event candidates and  $\met>30\GeV$ for 
the others).
Contamination in the control
region from signal and background processes considered in the analysis is 
predicted by MC simulations and is subtracted.
A scale factor, the ratio between the number of events predicted 
in the signal and control regions, is determined using MC simulations 
and is used to extrapolate the $\Zg$+jets event rate from the 
control region measured in data into the signal region.
Although the predictions from MC calculations agree with the data-driven 
estimates, the estimates have smaller uncertainties.

Non-$Z$\ boson backgrounds mainly come from $W$+jets, \ttbar\ 
production with a single lepton in the final state and single top
production. Such background events contain non-prompt leptons (e.g.\ leptons coming from 
$b$-hadron decays) or misidentified leptons arising from jets
(e.g.\ lighter hadron decays with a leading $\pi^0$ decaying to photons). 
The term ``fake lepton''
refers to both misidentified and non-prompt lepton candidates. 

The yield of background events with two well-identified lepton candidates
that contain at least one fake lepton
is estimated from data using a matrix method~\cite{top2010}.
From data control regions the probability for single loose leptons to pass the full identification cuts (tight leptons) is measured. 
A loose lepton refers to a lepton candidate that passes looser isolation criteria.
The control regions are selected such that either dominantly real or fake leptons are selected by the looser cuts.
The probability for real leptons is measured from the $\Zee$\ and
$\Zmm$\ control regions. 
The probability for
fake leptons is measured in a data sample dominated by dijet production
with events containing one loose lepton candidate and having low $\met$.
These probabilities enter a matrix that relates the numbers of observed 
dilepton candidate events with every combination of loose or tight leptons with
the numbers of events from the sources of either real leptons or objects
that might result in a fake lepton candidate.
The matrix is inverted in order to estimate the real and fake content of the
observed event sample.

In the lepton+track channels, the largest source of non-$Z$ boson 
backgrounds are events with a fake track lepton candidate.
This background rate is determined from 
a $\gamma$+jets data sample selected with photon triggers.
The fake rate is applied to a second sample enriched in $W$+jets events with 
exactly one lepton and no track leptons but using the same kinematic cuts 
as for the signal sample. 
In this second sample the fake probabilities are summed over all jets in all events
and the fake rates are calculated as a function of the jet multiplicity. 

The contributions from other electroweak background
processes with two real leptons (other EW), such as single top, $Z \to \tau\tau$, $WW$, $ZZ$ and $WZ$
production are estimated from Monte-Carlo simulations and 
found to be relatively small.
The numbers of
background events estimated with each method are
included in Table~\ref{t:signal}.

\newcommand{\DYZeeNJetsTwoJet}{1.1 $\pm$ 0.5}
\newcommand{\DYZmmNJetsTwoJet}{3.5 $\pm$ 1.4}
\newcommand{\ZtteeNJetsTwoJet}{0.4 $\pm$ 0.3}
\newcommand{\ZttmmNJetsTwoJet}{1.2 $\pm$ 0.6}
\newcommand{\ZttemNJetsTwoJet}{3.0 $\pm$ 1.3}
\newcommand{\FakeWeeNJetsTwoJet}{1.0 $\pm$ 0.9}
\newcommand{\FakeWmmNJetsTwoJet}{0.4 $\pm$ 0.5}
\newcommand{\FakeWemNJetsTwoJet}{1.9 $\pm$ 1.7}
\newcommand{\singletopeeNJetsTwoJet}{0.6 $\pm$ 0.1}
\newcommand{\singletopmmNJetsTwoJet}{1.2 $\pm$ 0.2}
\newcommand{\singletopemNJetsTwoJet}{2.4 $\pm$ 0.3}
\newcommand{\dibosoneeNJetsTwoJet}{0.5 $\pm$ 0.1}
\newcommand{\dibosonmmNJetsTwoJet}{0.9 $\pm$ 0.1}
\newcommand{\dibosonemNJetsTwoJet}{$2.0^{ +0.3}_{ -0.2}$}
\newcommand{\ttbareeNJetsTwoJet}{10.9 $\pm$ 1.2}
\newcommand{\ttbarmmNJetsTwoJet}{19.4 $\pm$ 1.5}
\newcommand{\ttbaremNJetsTwoJet}{45.7 $\pm$ 3.7}
\newcommand{\DataeeNJetsTwoJet}{17}
\newcommand{\DatammNJetsTwoJet}{30}
\newcommand{\DataemNJetsTwoJet}{57}
\newcommand{\TotalNonttbaree}{3.6 $\pm$ 1.2}
\newcommand{\TotalNonttbarmm}{7.2 $\pm$ 1.6}
\newcommand{\TotalNonttbarem}{9.4 $\pm$ 2.5}
\newcommand{\TotalExpectedee}{14.5 $\pm$ 1.7}
\newcommand{\TotalExpectedmm}{26.6 $\pm$ 2.1}
\newcommand{\TotalExpectedem}{55.1 $\pm$ 4.4}

\begin{table*}[t]
\centering
\begin{tabular}{|l|c|c|c|c|c||c|c|c|} \hline
  \multirow{2}{*}{} & \multicolumn{5}{c||}{Untagged} & \multicolumn{3}{c|}{Tagged}   \\ \hline
                          & $ee$                    & $\mu\mu$                & $e\mu$                  & $e$TL                & $\mu$TL      & $ee$         &  $\mu\mu$      &  $e\mu$          \\ [0.2ex] \hline
                          &                         &                         &                         &                      &              &                      &                       &                         \\ [-1.9ex]
                          $\Zg \to ee$/$\mu\mu$                & \DYZeeNJetsTwoJet       & \DYZmmNJetsTwoJet       & -                       & $7.1 \pm 1.5$  & $2.2\pm0.9$  & $2.6 \pm 1.3 $       & $5.0\pm1.7$           & -                       \\ [0.3ex]
$\Zg \to \tau\tau$    & \ZtteeNJetsTwoJet       & \ZttmmNJetsTwoJet       & \ZttemNJetsTwoJet       & $1.9 \pm 1.0$  & $2.2\pm0.9$  & $0.2  \pm 0.1$       & $0.2 \pm 0.1$         & $0.8 \pm 0.4$           \\ [0.3ex]
Fake leptons & \FakeWeeNJetsTwoJet     & \FakeWmmNJetsTwoJet     & \FakeWemNJetsTwoJet     & $8.1\pm2.9$          & $8.2\pm 2.9$ & $0.5 \pm 0.5$        & $0.4  \pm 0.5$        & $0.2 \pm 1.1$           \\ [0.3ex]
Single Top                & \singletopeeNJetsTwoJet & \singletopmmNJetsTwoJet & \singletopemNJetsTwoJet & $0.5\pm 0.1$         & $0.6\pm0.1$  & $0.6  \pm 0.1 $      & $1.1 \pm 0.2$         & $1.8 \pm 0.3$           \\ [0.3ex]
Diboson                  & \dibosoneeNJetsTwoJet   & \dibosonmmNJetsTwoJet   & \dibosonemNJetsTwoJet   & $0.5\pm0.1$          & $0.4\pm0.1$  & $0.2  \pm 0.1$       & $0.2 \pm 0.0$         & $0.4 \pm 0.1$           \\ [0.3ex] \hline
                          &                         &                         &                         &                      &              &                      &                       &                         \\ [-1.9ex]
Total Background           & \TotalNonttbaree        & \TotalNonttbarmm        & \TotalNonttbarem        & $18.1 \pm  3.4 $     & $13.8\pm3.2$ & $4.1 \pm 1.4       $ & $6.9 \pm 1.8$         & $3.2 \pm 1.2$           \\ [0.3ex]
Predicted \ttbar\       & \ttbareeNJetsTwoJet     & \ttbarmmNJetsTwoJet     & \ttbaremNJetsTwoJet     & $10.2 \pm 1.3$       & $11.0\pm1.8$ & $11.1\pm1.4$ & $20.6^{+1.7}_{-2.2} $ & $38.9^{+ 3.5}_{- 4.4}$  \\ [0.3ex] \hline
                          &                         &                         &                         &                      &              &                      &                       &                         \\ [-1.9ex]
Total                 & \TotalExpectedee        & \TotalExpectedmm        & \TotalExpectedem        & $28.3\pm 3.6$ & $24.6\pm3.7$ & $15.2\pm2.0$ & $27.5^{+2.5}_{-2.9}$  & 42.1 $^{+ 3.7}_{- 4.6}$ \\ [0.3ex] \hline \hline
                          &                         &                         &                         &                      &              &                      &                       &                         \\ [-1.9ex]
Observed               & \DataeeNJetsTwoJet      & \DatammNJetsTwoJet      & \DataemNJetsTwoJet      & 29                   & 21           & 17                   & 32                    & 49                      \\ \hline
\end{tabular}
\caption
{Full breakdown of the expected {\ttbar} signal and background events
compared to the observed event yields for each dilepton channel.
For the expected number of events a \ttbar\ cross section of $165\pmasym{11}{16}$~pb~\cite{topxs,Langenfeld:2009tc} is used. 
All systematic uncertainties are included and correlations between
different background sources are taken into account.
The fake leptons category includes both misidentified and non-prompt
lepton candidates.
}
\label{t:signal}
\end{table*}

The modeled acceptances, efficiencies and data-driven background
estimation methods are validated by comparing Monte-Carlo predictions
with data in control regions that are depleted of $\ttbar$ events
but have similar kinematics.  
In particular, the \MET, $m_{\ell\ell}$ and jet multiplicity distributions
in a sample of $Z$~boson candidates 
defined by requiring $|m_{\ell\ell} - m_Z| < 10\GeV$ and low \MET\ 
are compared to 
MC predictions and are in good agreement with data.

The background contributions after requiring at least one $b$-tagged jet are 
determined using the
same techniques described above to evaluate the rate of 
the background sources before making the $b$-tag requirement. 
Measured light quark and gluon 
jet rejection factors~\cite{ATLAS-CONF-2011-089}\ are then applied
to estimate the number of background events that remain in the 
candidate sample.

\section{Systematic Uncertainties}
\label{s:objsyst}

The uncertainties due to MC simulation modeling of the lepton
trigger, lepton and track-lepton reconstruction and selection efficiencies are assessed using 
$Z\rightarrow ee$ and $Z\rightarrow\mu\mu$ candidate events found in the same
data sample used for the $\ttbar$ analyses before applying
$Z$~boson veto requirements.
Scale factors are applied to MC samples when calculating
acceptances to account for any observed differences in predicted and observed efficiencies.
The modeling of lepton momentum scale and resolution is studied using
the $m_{ll}$\ distributions of $\Zg$\ candidate events,
and the
simulation is adjusted accordingly.
The acceptance uncertainty from the lepton modeling is dominated
by the electron selection efficiency uncertainty.

The jet energy scale (JES) and its uncertainty are derived by
combining information from test-beam data, LHC collision data and
simulation~\cite{ATL-CONF-2011-032}. 
For the selected jets, the JES uncertainty varies 
in the range $2-8\%$
as a function of jet $\pT$ and $\eta$. The jet energy
resolution and jet reconstruction efficiency measured in data and in
simulation are compared and are in good agreement. The statistical uncertainties on the
comparisons,  10\%\ and $1-2$\%\ for the energy resolution and the efficiency,
respectively, are taken as systematic uncertainties associated with these effects.
The effect on the acceptance is dominated by the JES uncertainty.

The systematic uncertainty in the efficiency of the \JetProb\ tagging algorithm has been estimated to be $6$\%\ for $b$-quark jets, based on $b$-tagging calibration 
studies using inclusive lepton and multijet final 
states~\cite{ATLAS-CONF-2011-089}. 
The uncertainties on the tagging efficiencies for light and charm quarks are several times higher, but are not a large source of uncertainty due to the 
intrinsically high signal-to-background ratios in the dilepton final states.
The acceptance uncertainty due to $b$-tagging ranges from 4 to 6\%\ 
depending on the channel.

The uncertainty in the kinematic distribution of the \ttbar\ signal events
gives rise to systematic uncertainties in
the signal acceptance, with contributions from the choice of generator, the modeling of initial and final state
radiation (ISR/FSR) and the PDFs. 
The generator uncertainty is evaluated by comparing the  {\sc MC@NLO}\ MC predictions  with those of 
the {\sc Powheg}\ MC~\cite{powheg,Frixione:2007vw,Alioli:2010xd} interfaced to
both {\sc Herwig} or {\sc Pythia}~\cite{Sjostrand:2006za} shower models. 
The uncertainty due to ISR/FSR is evaluated using the {\sc AcerMC} generator~\cite{Acer}
interfaced to the {\sc Pythia}\ shower model, and by varying the parameters controlling ISR
and FSR in a range consistent with experimental data~\cite{CSCbook}.
Finally, the PDF uncertainty is evaluated using a range of
current PDF sets~\cite{top2010}.
The dominant uncertainty in this category of systematics is the modeling of ISR/FSR and generator choice.

For $\Zg$+jets background events the normalization uncertainty is modeled 
by separately considering events with a given jet multiplicity.
While the cross section in the 0-jet multiplicity sample has $4\%$ uncertainty, the extrapolation to each following jet multiplicity increases the uncertainty by
an additional $24\%$~\cite{Alwall:2008}.

Overall normalization uncertainties on the backgrounds from single top quark and diboson production
are taken to be 10~\%~\cite{Campbell:2004ch,Kidonakis:2010yq} and 5~\%~\cite{Campbell:2010ff}, respectively.

The systematic uncertainties from the background estimates employing
complementary samples include the statistical uncertainties as well as the systematic uncertainties arising from the objects and MC estimates that are used in the methods.
The uncertainty on the data-driven $\Zg$+jets estimation is included by varying the 
$\met$ cut in the control region by $\pm 5\GeV$.
An additional systematic uncertainty for the fake track-lepton estimate
is derived from the difference in the observed and
predicted number of fake events in control 
regions, defined as opposite sign events with zero or one jet without an $\HT$ cut or same sign-events with more than one jet. 
Both data-driven methods are limited primarily by the statistical uncertainty 
in the number of events in the respective control regions.

\begin{table*}[t]
\centering

\begin{tabular}{|l|r@{/\kern -0.055em}l|r@{/\kern -0.055em}l|r@{/\kern -0.055em}l|r@{/\kern -0.055em}l|r@{/\kern -0.055em}l|r@{/\kern -0.055em}l||r@{/\kern -0.055em}l|r@{/\kern -0.055em}l|r@{/\kern -0.055em}l|r@{/\kern -0.055em}l|}
\hline
\multirow{2}{*}{$\Delta \sigma / \sigma$ (\%) } &  \multicolumn{12}{c||}{Untagged} & \multicolumn{8}{c|}{Tagged} \\  \cline{2-21}
 & \multicolumn{2}{c}{$ee$} & \multicolumn{2}{c}{$\mu\mu$} & \multicolumn{2}{c}{$e\mu$} & \multicolumn{2}{c}{$e$TL} & \multicolumn{2}{c}{$\mu$TL} & \multicolumn{2}{c||}{comb.} 
 & \multicolumn{2}{c}{$ee$} & \multicolumn{2}{c}{$\mu\mu$} & \multicolumn{2}{c}{$e\mu$} & \multicolumn{2}{c|}{comb.} \\ \hline
            &\multicolumn{2}{|c|}{}&\multicolumn{2}{|c|}{}&\multicolumn{2}{|c|}{}&\multicolumn{2}{|c|}{}&\multicolumn{2}{|c|}{}&\multicolumn{2}{|c||}{}            
            &\multicolumn{2}{|c|}{}&\multicolumn{2}{|c|}{}&\multicolumn{2}{|c|}{}&\multicolumn{2}{|c|}{}            \\ [-1.9ex]
Statistics  & ${}^{+33}$ & ${}_{-29}$ & ${}^{+26}$ & ${}_{-23}$ & ${}^{+17}$ & ${}_{-15}$ & ${}^{+53}$ &  ${}_{-46}$ &  ${}^{+67}$ & ${}_{-58}$ & ${}^{+12}$ & ${}_{-11}$
            & ${}^{+35}$ & ${}_{-29}$ & ${}^{+24}$ & ${}_{-21}$ & ${}^{+16}$ & ${}_{-15}$ & ${}^{+12}$ &  ${}_{-11}$ \\ 
MC Stat.    & ${}^{+5}$  & ${}_{-1}$  & ${}^{+4}$  & ${}_{-3}$  & ${}^{+2}$  & ${}_{-1}$  & ${}^{+12}$ &  ${}_{-14}$ &  ${}^{+10}$ & ${}_{-12}$ & ${}^{+1}$  & ${}_{-2}$
            & ${}^{+3}$  & ${}_{-5}$  & ${}^{+4}$  & ${}_{-3}$  & ${}^{+1}$  & ${}_{-1}$  & ${}^{+1}$  &  ${}_{-2}$  \\ 
Lepton      & ${}^{+11}$ & ${}_{-5}$  & ${}^{+4}$  & ${}_{-0}$   & ${}^{+5}$  & ${}_{-4}$  & ${}^{+7}$  &  ${}_{-6}$  &  ${}^{+3}$  & ${}_{-4}$  & ${}^{+4}$  & ${}_{-3}$
            & ${}^{+9}$ & ${}_{-7}$   & ${}^{+2}$  & ${}_{-2}$  & ${}^{+5}$  & ${}_{-4}$  & ${}^{+4}$  &  ${}_{-3}$  \\ 
$e$TL/$\mu$TL   & \multicolumn{2}{c|}{-}  & \multicolumn{2}{c|}{-}  & \multicolumn{2}{c|}{-}  & ${}^{+3}$  &  ${}_{-2}$  &  ${}^{+4}$  & ${}_{-5}$  & ${}^{+1}$  & ${}_{-1}$
            & \multicolumn{2}{c|}{-}  & \multicolumn{2}{c|}{-}  & \multicolumn{2}{c|}{-}  & \multicolumn{2}{c|}{-}   \\ 
Jet/$\met$  & ${}^{+5}$  & ${}_{-4}$  & ${}^{+4}$  & ${}_{-3}$  & ${}^{+3}$  & ${}_{-3}$  & ${}^{+13}$ &  ${}_{-11}$ &  ${}^{+12}$  & ${}_{-7}$  & ${}^{+3}$  & ${}_{-4}$
            & ${}^{+5}$  & ${}_{-4}$  & ${}^{+6}$  & ${}_{-4}$  & ${}^{+2}$  & ${}_{-2}$  & ${}^{+4}$  &  ${}_{-3}$  \\ 
$\Zg$+jets  & ${}^{+4}$  & ${}_{-4}$  & ${}^{+4}$  & ${}_{-3}$  & \multicolumn{2}{c|}{-}  & ${}^{+0}$  &  ${}_{-5}$  &  ${}^{+4}$  & ${}_{-6}$  & ${}^{+1}$  & ${}_{-1}$
            & ${}^{+6}$  & ${}_{-8}$  & ${}^{+5}$  & ${}_{-6}$  & \multicolumn{2}{c|}{-}  & ${}^{+2}$  &  ${}_{-2}$  \\ 
Fake        & ${}^{+8}$  & ${}_{-6}$  & ${}^{+3}$  & ${}_{-1}$  & ${}^{+3}$  & ${}_{-4}$  & ${}^{+25}$ &  ${}_{-27}$ &  ${}^{+39}$ & ${}_{-41}$ & ${}^{+3}$  & ${}_{-3}$
            & ${}^{+2}$  & ${}_{-5}$  & ${}^{+2}$  & ${}_{-2}$  & ${}^{+3}$  & ${}_{-2}$  & ${}^{+1}$  &  ${}_{-2}$  \\ 
Generator   & ${}^{+6}$  & ${}_{-4}$  & ${}^{+5}$  & ${}_{-6}$  & ${}^{+4}$  & ${}_{-4}$  & ${}^{+16}$ &  ${}_{-11}$ &  ${}^{+17}$ & ${}_{-17}$  & ${}^{+5}$  & ${}_{-5}$
            & ${}^{+10}$ & ${}_{-8}$  & ${}^{+7}$  & ${}_{-5}$  & ${}^{+5}$  & ${}_{-4}$  & ${}^{+6}$  &  ${}_{-4}$  \\ 
$b$-tagging & \multicolumn{2}{c|}{-}  & \multicolumn{2}{c|}{-}  & \multicolumn{2}{c|}{-}  & \multicolumn{2}{c|}{-}   & \multicolumn{2}{c|}{-}   & \multicolumn{2}{c||}{-}
            & ${}^{+3}$ & ${}_{-4}$  & ${}^{+5}$ & ${}_{-3}$  & ${}^{+5}$  & ${}_{-4}$  & ${}^{+5}$  &  ${}_{-4}$  \\   
Luminosity  & ${}^{+4}$ & ${}_{-4}$  & ${}^{+4}$ & ${}_{-4}$  & ${}^{+4}$  & ${}_{-4}$  & ${}^{+4}$  &  ${}_{-4}$    & ${}^{+5}$   & ${}_{-5}$  & ${}^{+4}$  & ${}_{-4}$ 
            & ${}^{+4}$ & ${}_{-3}$  & ${}^{+4}$ & ${}_{-3}$  & ${}^{+4}$  & ${}_{-4}$  & ${}^{+3}$  &  ${}_{-4}$  \\  [+1.05 ex] \hline  
\end{tabular}

\caption{The \ttbar{} cross section uncertainties. 
These include the uncertainties from the data 
and MC statistics, the uncertainties related to the object selection 
(grouped in lepton, track lepton $e$TL/$\mu$TL, jet/$\met$ and $b$-tagging 
uncertainties), the background estimation methods ($\Zg$+jets and fakes), the uncertainties 
on the simulated samples (generator) and the luminosity uncertainty.
}
\label{t:xssys}
\end{table*}

\section{Cross Section Measurement}
\label{s:nobtagging}

The expected and measured numbers of events in the signal region after
applying all selection cuts for each of the individual
dilepton channels are shown in Table~\ref{t:signal}.
A total of 154 candidate events is observed for the 
analysis without $b$-tagging, 104 events in the well-identified dilepton 
channels 
and 50 events in the lepton+track channels.
A total of 98 candidate events are observed in the analysis using
$b$-tagging,
with 84 events in common with the untagged analysis.

\begin{figure*}
  \centering
 \includegraphics[width=0.245\textwidth]{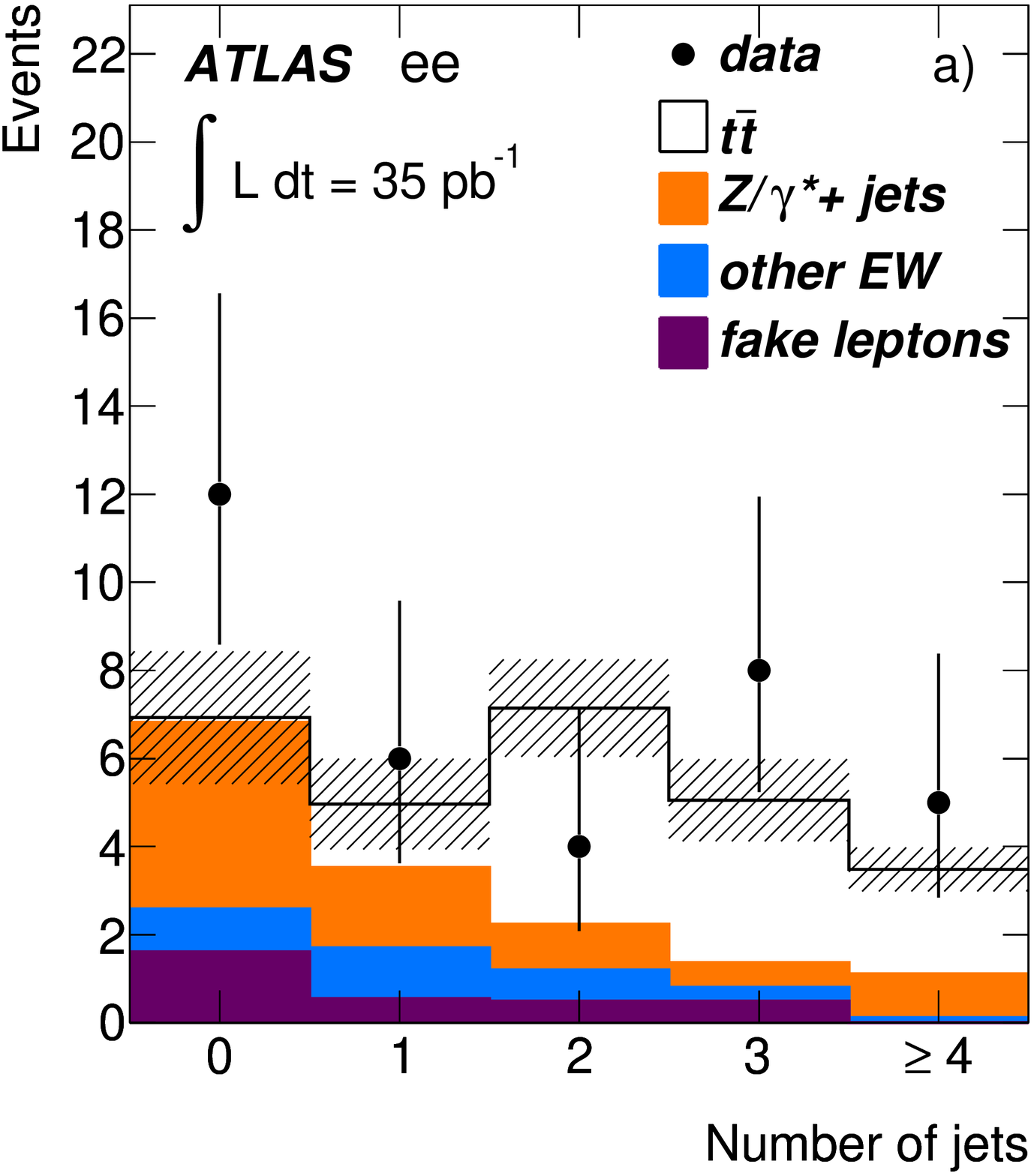} 
 \includegraphics[width=0.245\textwidth]{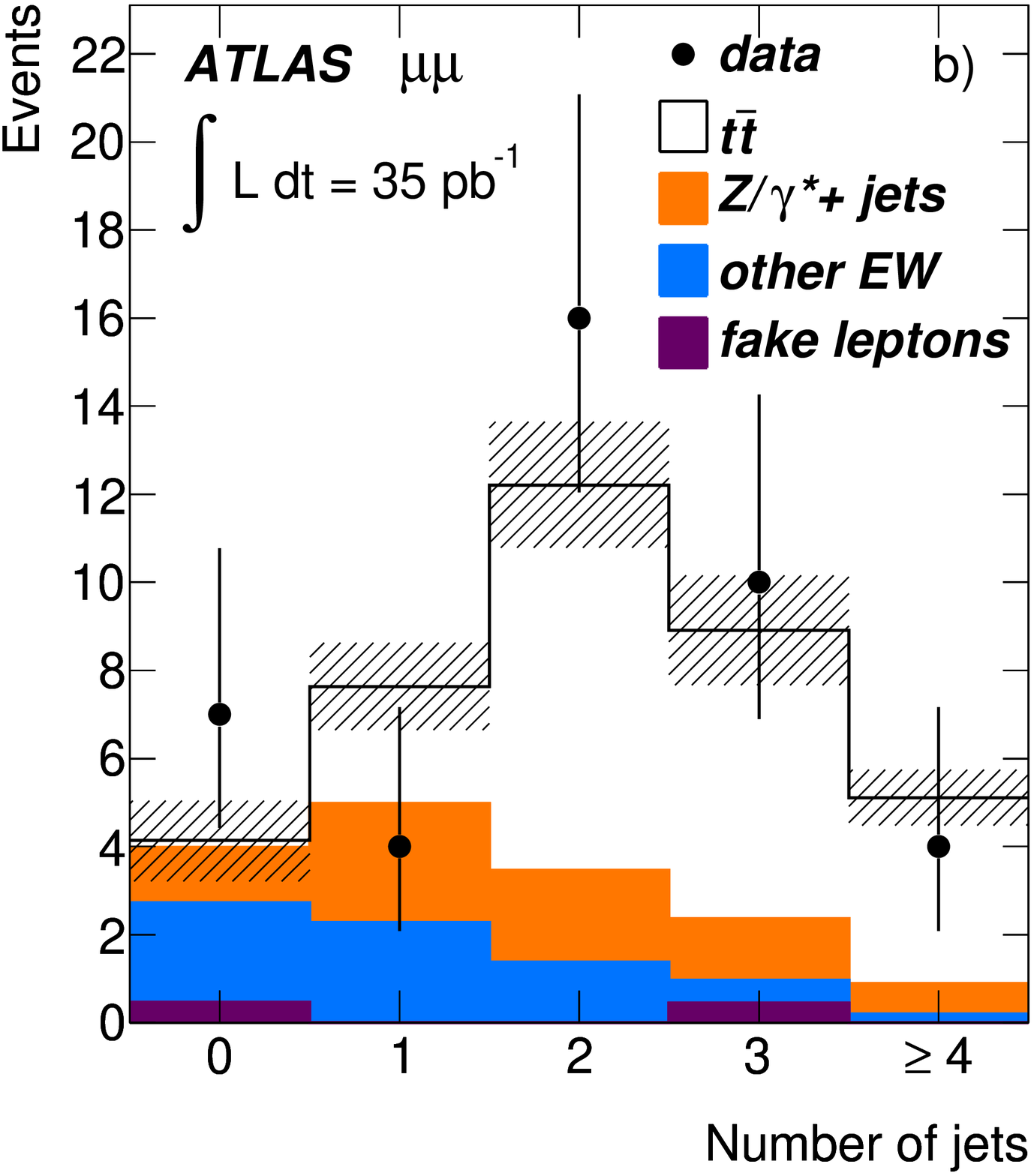} 
 \includegraphics[width=0.245\textwidth]{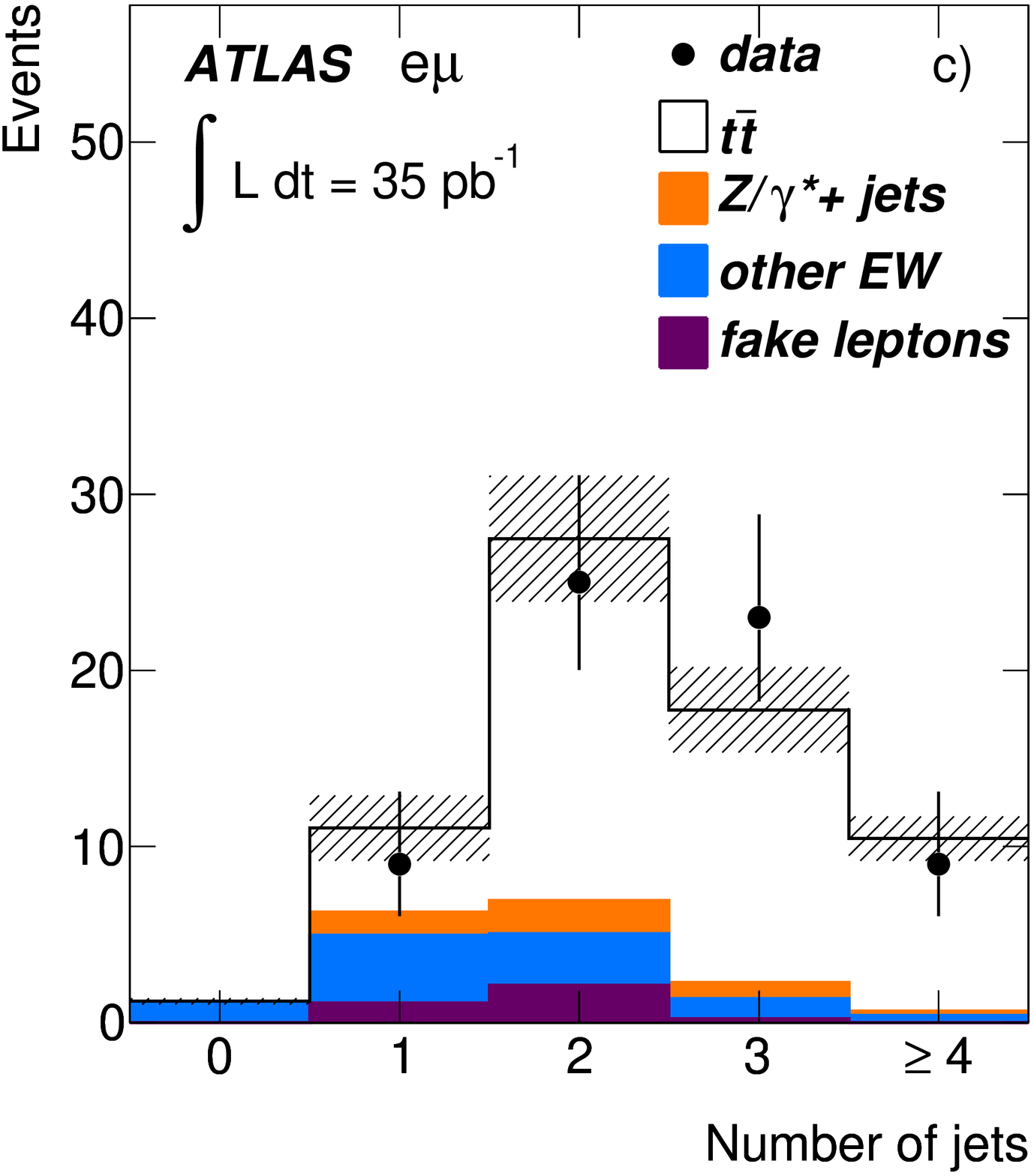} 
 \includegraphics[width=0.245\textwidth]{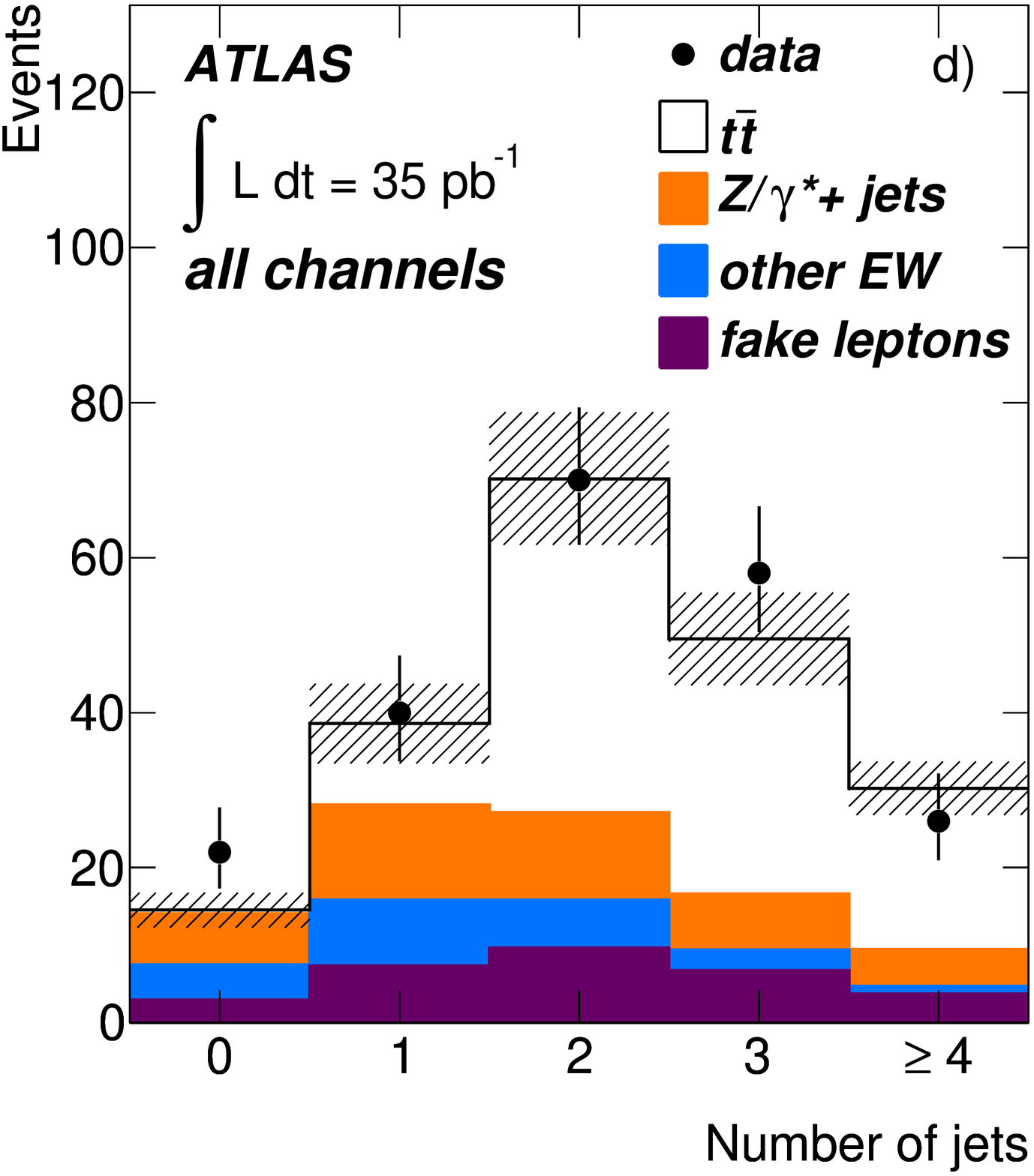}
  \caption{ Jet multiplicity distributions for the signal region omitting the 
$N_\mathrm{jets}\ge 2$ requirement in (a) the $ee$ channel, 
(b) the $\mu\mu$ channel, 
(c) the $e\mu$ channel and 
(d) all five channels combined.
The fake lepton contribution in (d) is the sum of the fake track-lepton and the fake lepton contribution.
Contributions from diboson and single top events are summarized as `other EW'.
The uncertainty on the data points are statistical uncertainties only, whereas the uncertainty bands include statistical and systematic uncertainies.
}
  \label{f:ll_njets}
\end{figure*}

\begin{figure*}
  \centering
 \includegraphics[width=0.325\textwidth]{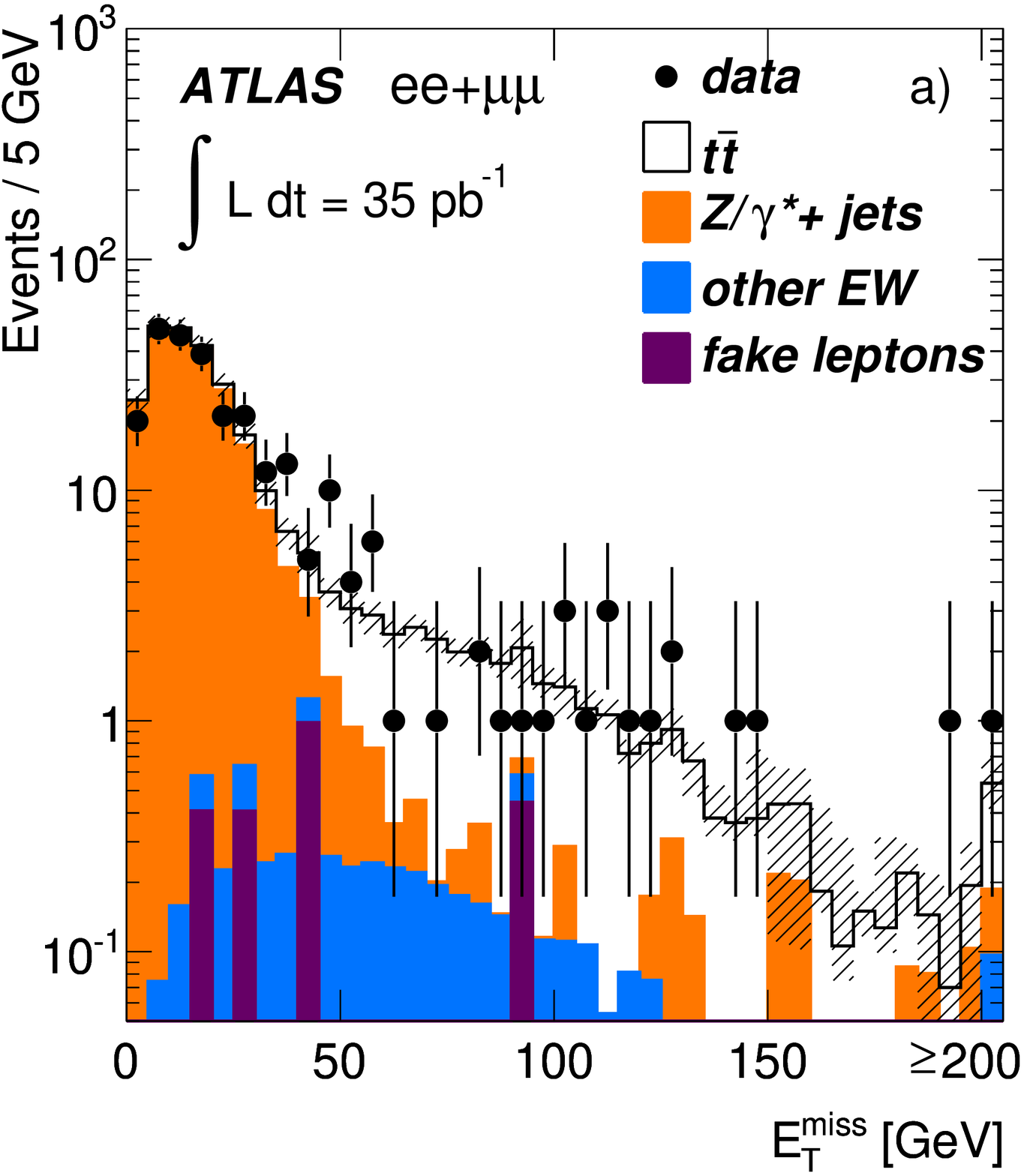} 
 \includegraphics[width=0.31\textwidth]{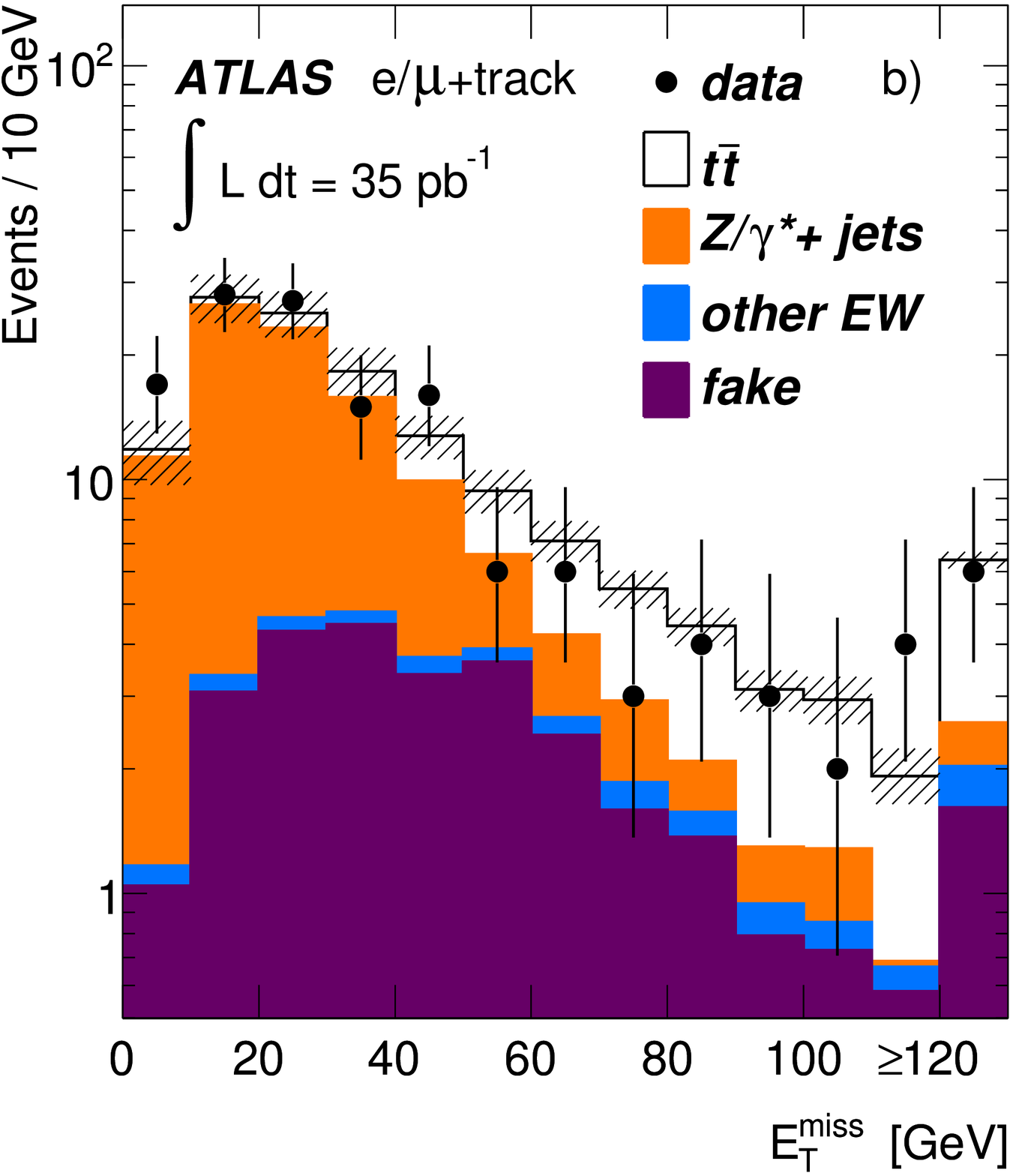} 
 \includegraphics[width=0.325\textwidth]{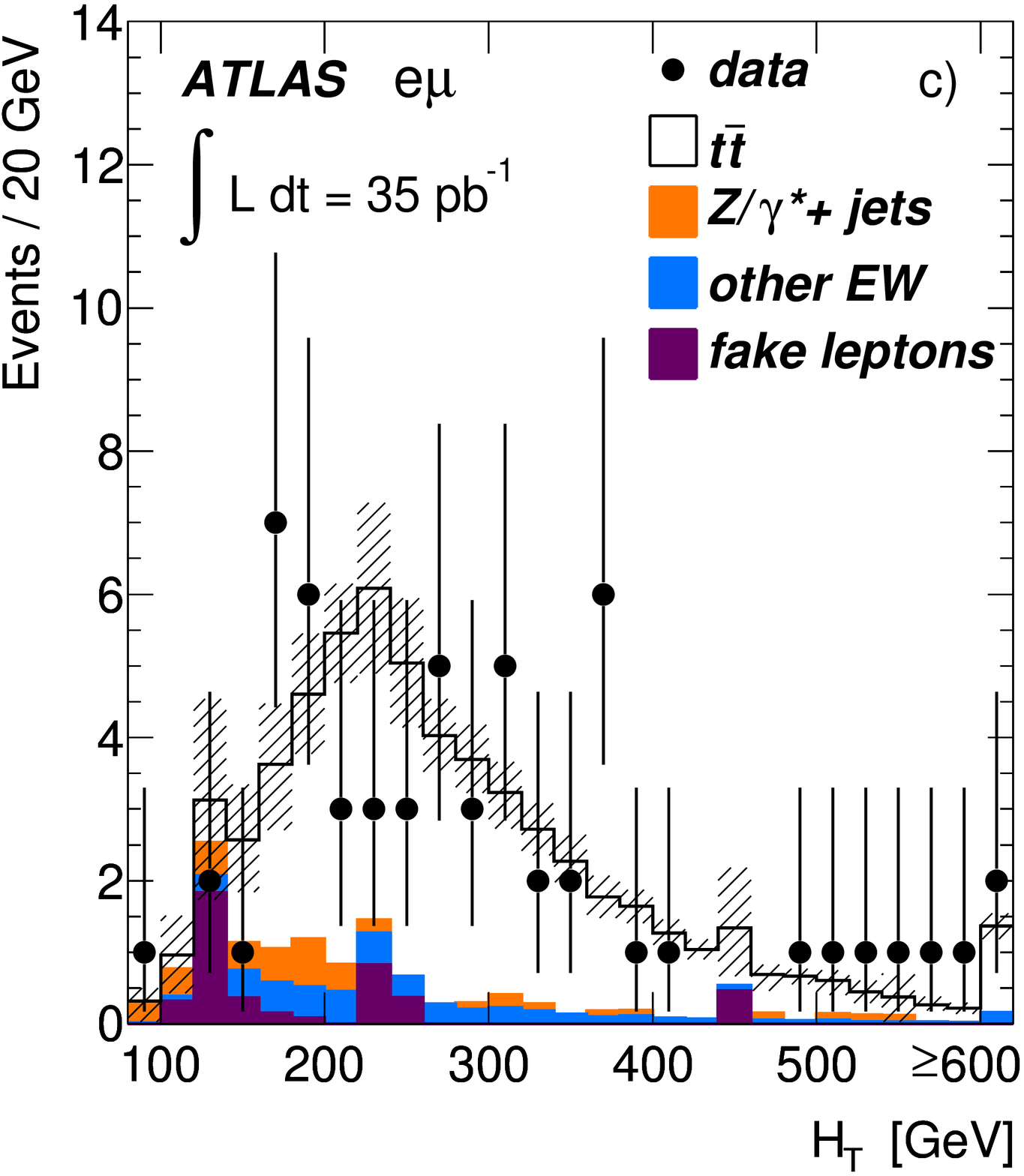} 
  \caption{
The $\met$ distribution in the signal region 
without the
$\MET>40\GeV$ requirement
(a) for the $ee$ and $\mu\mu$  channels and (b) for the lepton+track channels.  Fake denotes the contribution from fake track-leptons.
The $\HT$\ distribution in the signal region for the $e\mu$\ channel 
is shown in (c) 
without the $\HT>130\GeV$ requirement.  
Contributions from diboson and single top events are summarized as `other EW'.
In all figures the last bin contains the overflow.
The uncertainty on the data points are statistical uncertainties only, whereas the uncertainty bands include statistical and systematic uncertainies.
}
  \label{f:ll_met_ht}
\end{figure*}

\begin{figure*}
 \centering
 \includegraphics[width=0.325\textwidth]{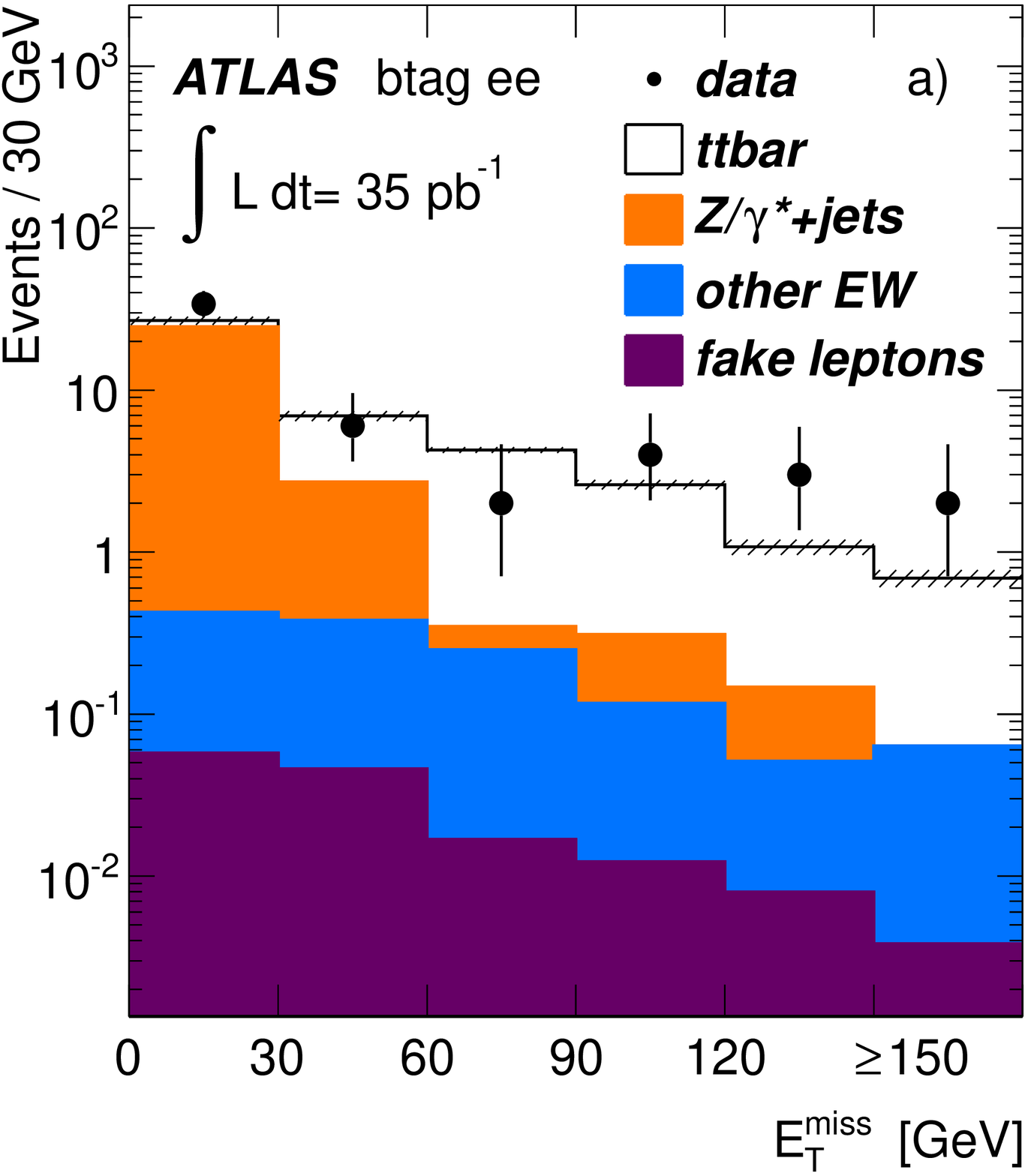}
 \includegraphics[width=0.325\textwidth]{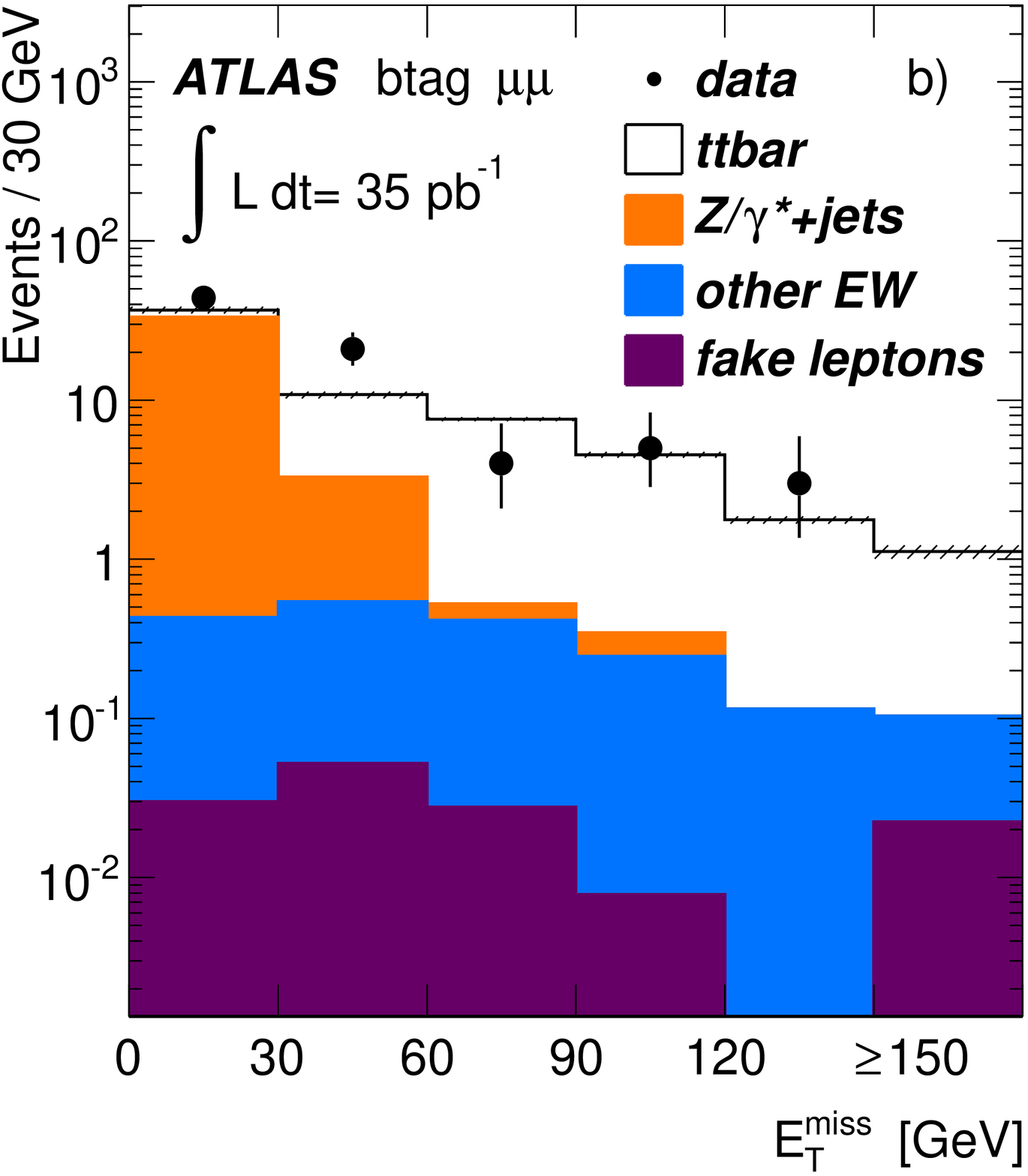}
 \includegraphics[width=0.325\textwidth]{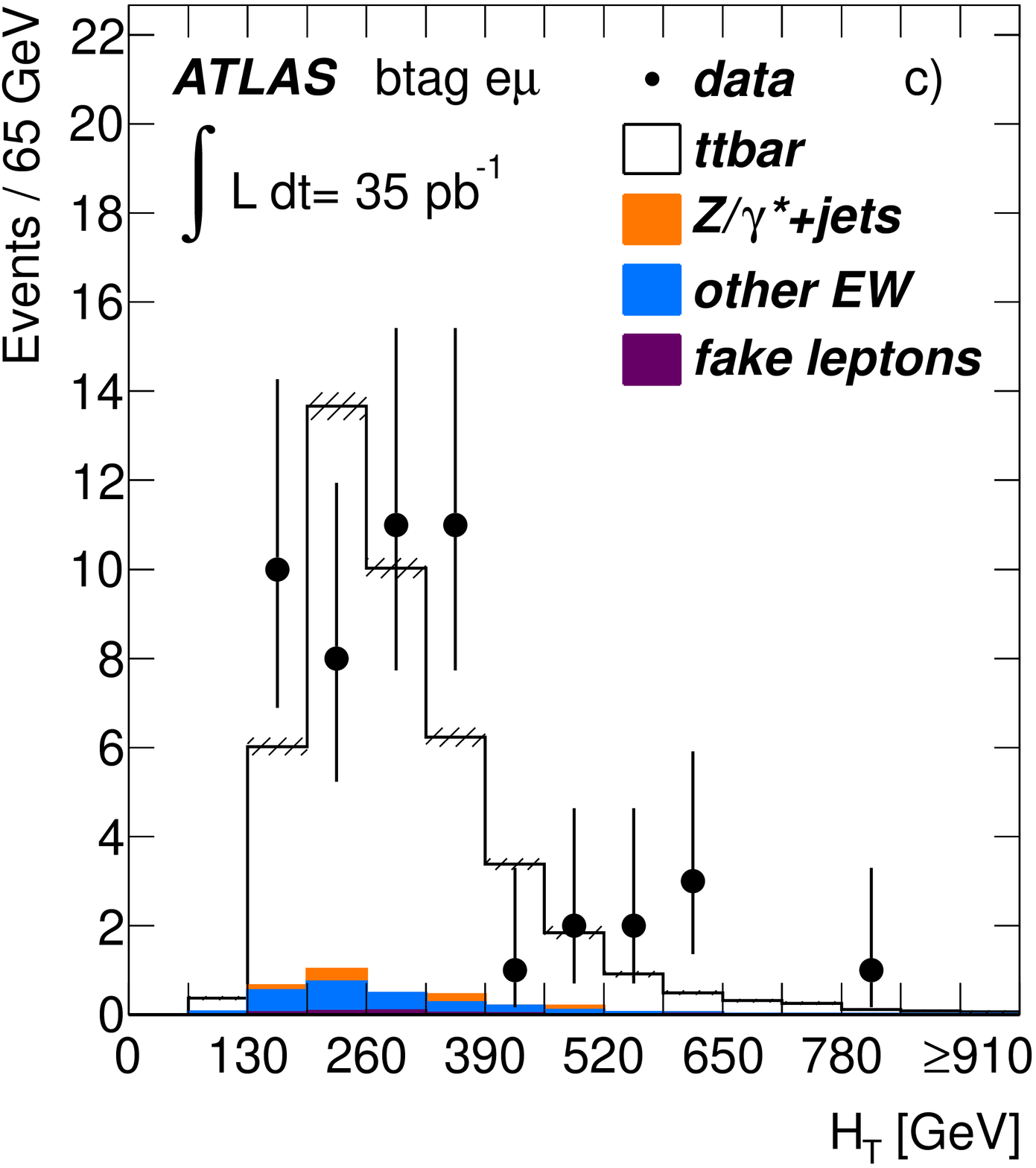}
 \caption{The $\MET$ distributions for the (a) $ee$ and (b) $\mu\mu$ channels omitting the $\met$ requirement, and (c) the $\HT$ distribution for the $e\mu$ channel omitting the $\HT$\ requirement, in each case after $b$-tagging has been applied.  Contributions from diboson and single top events are summarized as `other EW'.
The last bin in all figures contains the overflow. 
The uncertainty on the data points are statistical uncertainties only, whereas the uncertainty bands include statistical and systematic uncertainies.
}
\label{f:btaggingMEtHt}   
\end{figure*}

In Fig.~\ref{f:ll_njets} the distributions of the jet multiplicity 
are shown for the $ee$, $\mu\mu$\ and $e\mu$ channels and the sum of 
all five
channels together with the expectation for $\lumitot$. 
The distributions of $\met$ for the sum of the $ee$ and $\mu\mu$
channels, the sum of the track-lepton channels and of $\HT$ for the $e\mu$ channel are shown in
Fig.~\ref{f:ll_met_ht} and for the $b$-tag analysis in 
Fig.~\ref{f:btaggingMEtHt}.
All requirements are applied except on the variable whose distribution
is shown in the figure.

The dominant background in
the $ee$\ and $\mu\mu$\ channels is $\Zg +$jets production. 
The next largest background are events with fake leptons. From simulation it is found that this is mainly $W$+jets production with an
additional lepton candidate (mostly from $b$-quark decays).

The cross section results are obtained with a likelihood fit~\cite{ATL-CONF-2011-040} in which the 
probability of observing a number of
signal and background events, $N^\mathrm{obs}_i$, in each 
channel $i$ is modeled by a 
Poisson distribution, $\mathcal{P}$, 
given an expected number of events, $N^\mathrm{exp}_{i,\mathrm{tot}}$. 
The integrated luminosity, $L$, is modeled with a Gaussian distribution, $\mathcal{G}$,
about its central value, $L_0$.
The variation in $N^\mathrm{exp}_{i,\mathrm{tot}}$\ due to each 
systematic source $j$ is modeled with a 
Gaussian distribution, $\mathcal{G}_j$, for the associated nuisance 
parameter $\alpha_j$, where
$\alpha_j = \pm 1$ represents the $\pm$ 1 standard deviation variation of 
the systematic source.
The cross section, $\sigma_\mathrm{sig}$, is left as a free parameter in the fit of the likelihood function:

\begin{eqnarray*}\label{Eq:likelihood}
\mathcal{L}(\sigma_\mathrm{sig}, L,\vec{\alpha}) & = & \prod_{i\in \{ \rm channel\} } \mathcal{P}
\left (N^\mathrm{obs}_i \, | \, N^\mathrm{exp}_{i, \mathrm{tot}}(\vec{\alpha}) \right) \\ & \times&  
\mathcal{G}(L_0 | L, \sigma_L)\,  \times \prod_{j\in \rm syst} \mathcal{G}_j(0| \alpha_{j},1)  \, . \nonumber
\end{eqnarray*}

The cross section is inferred from the profile likelihood ratio
$\lambda(\sigma_\mathrm{sig}) = \mathcal{L}(\sigma_\mathrm{sig}, \hat{\hat{L}},
\hat{\hat{\vec{\alpha}}}) / \mathcal{L}(\hat\sigma_\mathrm{sig}, \hat L, 
\hat{\vec{\alpha}})$, where a single circumflex represents the maximum
likelihood estimate (MLE) of the parameter and the double circumflex
represents the conditional MLE for a given $\sigma_\mathrm{sig}$.  Ensembles
of pseudo-data are generated for $N^\mathrm{obs}_i$ and the resulting
estimate of $\hat \sigma_\mathrm{sig}$ is confirmed to be unbiased.
Additionally, the variance of $\hat \sigma_\mathrm{sig}$ is found to be
consistent with the curvature of the profile likelihood at its
minimum
and with the mean square spread observed in the ensemble tests.
Table~\ref{t:xssys}\ lists the uncertainties for each
contribution from the data and MC statistics, the uncertainties related to the object selection (grouped in lepton, track-lepton, jet/$\met$\  and $b$-jet uncertainties), the background estimation methods 
and the uncertainties on the simulated samples.
The variation of the cross section due to the luminosity uncertainty is obtained by repeating the likelihood minimization while fixing the luminosity to the nomimal value $\pm 1$ standard deviation.
For the final result the luminosity uncertainty is the difference of the total uncertainties for the likelhood function with and without the luminosity term.
Table~\ref{t:dilxs} summarizes the cross
sections extracted from the profile likelihood ratio for the
individual channels and for the combination of all channels for the analysis with and without a $b$-tagging requirement, respectively.
  
\section{Results}
\label{s:summary}

\begin{table}[t]
\centering 
\begin{tabular}{|c|c|c|} \hline
Channel   & $\sigma_{\ttbar}$ (pb)                                      & $b$-tag $\sigma_{\ttbar}$ (pb)                                 \\ [1.2ex] \hline
          &                                                             &                                                                \\ [-1.5ex]
$e e$     & $ 202 {\,}^{+67}_{-57} {\,}^{+30}_{-26}$    & $ 190 {\,}^{+66}_{-56} {\,}^{+36}_{-28}$       \\ [+1.2ex]
$\mu \mu$ & $ 192 {\,}^{+49}_{-44} {\,}^{+17}_{-15}$    & $ 200 {\,}^{+48}_{-42} {\,}^{+26}_{-20}$       \\ [+1.2ex]
$e \mu$   & $ 172 {\,}\pm 27 {\,}\pm 13            $    & $ 193 {\,}^{+31}_{-28} {\,}^{+18}_{-13}$        \\ [+1.2ex]\hline
          &                                                             &                                                                \\ [-1.5ex]
$e$TL     & $ 175 {\,}^{+92}_{-81} {\,}^{+65}_{-59}$    & $-$                                                                \\ [+1.2ex]
$\mu$TL   & $ 110 {\,}^{+74}_{-64} {\,}^{+56}_{-49}$    & $-$                                                                \\ [+1.2ex]\hline
          &                                                             &                                                                 \\ [-1.5ex]
Combined  & 171 $\pm 22$ $\pm 15$                       & 194 $\pm 23$ $^{ +18 }_{ -14 }  $              \\ [1ex] \hline
\end{tabular}
\caption{
Measured cross sections in each dilepton channel, and the 
combination of the untagged and tagged channels with their statistical and systematic uncertainties. The luminosity uncertainty is not included here.
}
\label{t:dilxs}
\end{table}

The top quark pair production cross section is measured using  events
selected by requiring two oppositely-charged lepton candidates, at least two 
additional jets and missing transverse energy.
The result is
\[
\sigmattbar=\mathrm{}\xsectot \xsecstat \mathrm{(stat.)}\xsecsyst \mathrm{(syst.)} \xseclumi \mathrm{(lum.)}~\rm pb. 
\]
A measurement made requiring at least one of the jets to be identified as 
a $b$-quark jet results in
\[
\sigmattbar=\mathrm{}\xsecbtot \xsecbstat  \mathrm{(stat.)}\xsecbsyst \mathrm{(syst.)} \xsecblumi  \mathrm{(lum.)}~\rm pb. 
\]
The two measurements agree with each other, taking into account that from all events 14\% (tagged analysis) and 45\% (untagged analysis) of the events are uncorrelated, and that the $b$-tagging systematic uncertainty is also uncorrelated.
The agreement 
confirms that the candidate events are consistent with arising from top quark pair production.

The measured cross sections are in good agreement with 
a similar measurement performed by the CMS collaboration~\cite{Chatrchyan:2011nb}, 
ATLAS measurements made in the complementary lepton+jets channels~\cite{ljpaper}\
and  
the SM prediction of $165\pmasym{11}{16}$~pb.

\section{Acknowledgements}

We thank CERN for the very successful operation of the LHC, as well as the
support staff from our institutions without whom ATLAS could not be
operated efficiently.

We acknowledge the support of ANPCyT, Argentina; YerPhI, Armenia; ARC,
Australia; BMWF, Austria; ANAS, Azerbaijan; SSTC, Belarus; CNPq and FAPESP,
Brazil; NSERC, NRC and CFI, Canada; CERN; CONICYT, Chile; CAS, MOST and
NSFC, China; COLCIENCIAS, Colombia; MSMT CR, MPO CR and VSC CR, Czech
Republic; DNRF, DNSRC and Lundbeck Foundation, Denmark; ARTEMIS, European
Union; IN2P3-CNRS, CEA-DSM/IRFU, France; GNAS, Georgia; BMBF, DFG, HGF, MPG
and AvH Foundation, Germany; GSRT, Greece; ISF, MINERVA, GIF, DIP and
Benoziyo Center, Israel; INFN, Italy; MEXT and JSPS, Japan; CNRST, Morocco;
FOM and NWO, Netherlands; RCN, Norway; MNiSW, Poland; GRICES and FCT,
Portugal; MERYS (MECTS), Romania; MES of Russia and ROSATOM, Russian
Federation; JINR; MSTD, Serbia; MSSR, Slovakia; ARRS and MVZT, Slovenia;
DST/NRF, South Africa; MICINN, Spain; SRC and Wallenberg Foundation,
Sweden; SER, SNSF and Cantons of Bern and Geneva, Switzerland; NSC, Taiwan;
TAEK, Turkey; STFC, the Royal Society and Leverhulme Trust, United Kingdom;
DOE and NSF, United States of America.

The crucial computing support from all WLCG partners is acknowledged
gratefully, in particular from CERN and the ATLAS Tier-1 facilities at
TRIUMF (Canada), NDGF (Denmark, Norway, Sweden), CC-IN2P3 (France),
KIT/GridKA (Germany), INFN-CNAF (Italy), NL-T1 (Netherlands), PIC (Spain),
ASGC (Taiwan), RAL (UK) and BNL (USA) and in the Tier-2 facilities
worldwide.

\bibliographystyle{atlasnote} 
\section*{References}
\bibliography{DileptonSigmaPLB}

\clearpage

\onecolumn
\input{atlas_authlist97final}

\end{document}

%% file: atlas_authlist97final.tex
% ATLAS Collaboration author list for 06-JUN-2011
% Data extracted on 26-Jun-2011 for paperid 97
%\documentclass[11pt]{article}
%\usepackage{a4wide}\begin{document}
\begin{flushleft}
{\Large The ATLAS Collaboration}

\bigskip

G.~Aad$^{\rm 48}$,
B.~Abbott$^{\rm 111}$,
J.~Abdallah$^{\rm 11}$,
A.A.~Abdelalim$^{\rm 49}$,
A.~Abdesselam$^{\rm 118}$,
O.~Abdinov$^{\rm 10}$,
B.~Abi$^{\rm 112}$,
M.~Abolins$^{\rm 88}$,
H.~Abramowicz$^{\rm 153}$,
H.~Abreu$^{\rm 115}$
% ...
\bigskip

% ...
$^{10}$ Institute of Physics, Azerbaijan Academy of Sciences, Baku, Azerbaijan\\
$^{11}$ Institut de F\'isica d'Altes Energies and Universitat Aut\`onoma  de Barcelona and ICREA, Barcelona, Spain\\
% ...
$^{48}$ Fakult\"{a}t f\"{u}r Mathematik und Physik, Albert-Ludwigs-Universit\"{a}t, Freiburg i.Br., Germany\\
$^{49}$ Section de Physique, Universit\'e de Gen\`eve, Geneva, Switzerland\\
% ...
$^{88}$ Department of Physics and Astronomy, Michigan State University, East Lansing MI, United States of America\\
% ...
$^{111}$ Homer L. Dodge Department of Physics and Astronomy, University of Oklahoma, Norman OK, United States of America\\
$^{112}$ Department of Physics, Oklahoma State University, Stillwater OK, United States of America\\
% ...
$^{115}$ LAL, Univ. Paris-Sud and CNRS/IN2P3, Orsay, France\\
% ...
$^{118}$ Department of Physics, Oxford University, Oxford, United Kingdom\\
% ...
$^{153}$ Raymond and Beverly Sackler School of Physics and Astronomy, Tel Aviv University, Tel Aviv, Israel\\
% ...
\end{flushleft}

%\end{document}